%% file: template_isit23.tex
\documentclass[conference,a4paper]{IEEEtran}

\usepackage[utf8]{inputenc} 
\usepackage[T1]{fontenc}
\usepackage{url}              % provides \url{...}
\usepackage{cite}             % improves presentation of citations

\usepackage[cmex10]{amsmath}  % Use the [cmex10] option to ensure complicance
\usepackage{mleftright}       % fix to wrong spacing of \left-,
\mleftright                   % \middle- \right-commands 

\usepackage{graphicx}         % provides \includegraphics{...} to
\usepackage{booktabs}         % fixes poor spacing in tables and
\usepackage{mathtools}

\usepackage{blindtext, graphicx}{}
\usepackage{amsmath,amsfonts,amssymb,bbm}
\usepackage{amsthm}

\usepackage[T1]{fontenc}
\usepackage{tabularx}
\usepackage[table]{xcolor}
\usepackage{epstopdf}
\usepackage{epsfig}
\graphicspath{{images/}}
\usepackage{dsfont}
\usepackage{psfrag,bbm,graphicx}
\usepackage{comment,balance}
\usepackage{hyperref}
\usepackage{epstopdf}
\usepackage{epsfig}
\usepackage{multicol}
\usepackage{multirow, cite}
\usepackage{amsfonts}
\usepackage{color}

\usepackage[ruled,vlined]{algorithm2e}
\hypersetup{
	colorlinks=true,
	linkcolor=blue,
	filecolor=magenta,      
	urlcolor=cyan,
}

\mathtoolsset{showonlyrefs}
\usepackage{geometry}

\newtheorem{theorem}{Theorem}
\newcommand{\Mod}[1]{\ (\mathrm{mod}\ #1)}
\newtheorem{example}{Example}
\newtheorem{lemma}{Lemma}
\newtheorem{proposition}[theorem]{Proposition}

\newtheorem{remark}[theorem]{Remark:}
\newcommand{\remove}[1]{}

\begin{document}

\title{Multi-access Coded Caching \\ with  Linear Subpacketization} 
%\vspace{-0.5 in}
%%%%%%
%\author{%
%  \IEEEauthorblockN{Anonymous Authors}
%%  \IEEEauthorblockA{%
%%    Please do NOT provide authors' names and affiliations\\
%%    in the paper submitted for review, but keep this placeholder.\\
%%    ISIT23 follows a \textbf{double-blind reviewing policy}.}
%}
%\author {
%	% Authors
%	Srinivas Reddy Kota,
%	P. N. Karthik, and
%	Vincent Y. F. Tan
%}
%\affiliations {
%	% Affiliations
%	%    \textsuperscript{\rm 1} Affiliation 1\\
%	%    \textsuperscript{\rm 2} Affiliation 2\\
%	National University of Singapore\\ Emails: ksvr1532@gmail.com, karthik@nus.edu.sg, vtan@nus.edu.sg
%}

%%%%%% Please only add the author names and affiliations for the FINAL
%%%%%% version of the paper, but NOT for the paper submitted for review!
%
%%%%%
%%%%% Single author, or several authors with same affiliation:
% \author{%
%   \IEEEauthorblockN{Stefan M.~Moser}
%   \IEEEauthorblockA{ETH Zürich\\
%                     8092 Zürich, Switzerland\\
%                     moser@isi.ee.ethz.ch}
%                   }
%
%%%%%
%%%%% Several authors with up to three affiliations:
 \author{%
   \IEEEauthorblockN{Srinivas Reddy Kota}
   \IEEEauthorblockA{Indian Institute of Technology Madras\\
                     ksreddy@ee.iitm.ac.in}
   \and
   \IEEEauthorblockN{Nikhil Karamchandani}
   \IEEEauthorblockA{Indian Institute of Technology Bombay\\
   	nikhilk@ee.iitb.ac.in
   	}
 }
 
%
%%%%%   
%%%%% Many authors with many affiliations:
% \author{%
%   \IEEEauthorblockN{Albus Dumbledore\IEEEauthorrefmark{1},
%                     Olympe Maxime\IEEEauthorrefmark{2},
%                     Stefan M.~Moser\IEEEauthorrefmark{3}\IEEEauthorrefmark{4},
%                     and Harry Potter\IEEEauthorrefmark{1}}
%   \IEEEauthorblockA{\IEEEauthorrefmark{1}%
%                     Hogwarts School of Witchcraft and Wizardry,
%                     1714 Hogsmeade, Scotland,
%                     \{dumbledore, potter\}@hogwarts.edu}
%   \IEEEauthorblockA{\IEEEauthorrefmark{2}%
%                     Beauxbatons Academy of Magic,
%                     1290 Pyrénées, France,
%                     maxime@beauxbatons.fr}
%   \IEEEauthorblockA{\IEEEauthorrefmark{3}%
%                     ETH Zürich, ISI (D-ITET), ETH Zentrum, 
%                     CH-8092 Zürich, Switzerland,
%                     moser@isi.ee.ethz.ch}
%   \IEEEauthorblockA{\IEEEauthorrefmark{4}%
%                     National Yang Ming Chiao Tung University (NYCU), 
%                     Hsinchu, Taiwan,
%                     moser@isi.ee.ethz.ch}
% }
%

\maketitle

%%%%%
%% Abstract: 
%% If your paper is eligible for the student paper award, please add
%% the comment "THIS PAPER IS ELIGIBLE FOR THE STUDENT PAPER
%% AWARD." as a first line in the abstract. 
%% For the final version of the accepted paper, please do not forget
%% to remove this comment!
%%

\begin{abstract}
We consider the multi-access coded caching problem, which contains a central server with $N$ files, $K$ caches with $M$ units of memory each and $K$ users where each one is connected to $L (\geq 1)$ consecutive caches, with a cyclic wrap-around. Caches are populated with content related to the files and each user then requests a file that has to be served via a broadcast message from the central server with the help of the caches. We aim to design placement and delivery policies for this setup that minimize the central servers' transmission rate while satisfying an additional linear sub-packetization constraint. We propose policies that satisfy this constraint and derive upper bounds on the achieved server transmission rate, which upon comparison with the literature establish the improvement provided by our results. To derive our results, we map the multi-access coded caching problem to variants of the well-known index coding problem. In this process, we also derive new bounds on the optimal transmission size for a `structured' index coding problem, which might be of independent interest.
%For the multi-access coded caching problem, we study a policy that satisfies linear sub-packetization restriction and derives the upper bounds on the optimal transmission rate of the central server under the restriction of linear sub-packetization. 
\end{abstract}

\input{introduction.tex}
\input{setting.tex}

\input{preliminaries.tex}

\input{Ourpolicy.tex}

\input{mainresults.tex}
\bibliographystyle{IEEEtran}
\balance
\bibliography{myref2}
\newpage
\appendices
\input{proofs.tex}

\end{document}

%% file: introduction.tex
\section{Introduction}\label{sec:introduction}{\let\thefootnote\relax\footnote{%This work was supported in part by the Bharti Centre for Communication. 
		The work of Srinivas Reddy Kota was supported in part by the INSPIRE Faculty Fellowship from the DST, Govt. of India. The work of Nikhil Karamchandani was supported in part by the SERB MATRICS grant. }}
The seminal work of Maddah-Ali and Niesen \cite{maddah2014fundamental} initiated an information-theoretic study of caching systems, which help reduce network congestion by offloading the data traffic from peak to off-peak hours. Following this, there has been a large amount of work studying various variants of this `coded caching' (CC) problem. In particular, motivated by wireless networks \cite{hachem2017codedmulti} introduced the multi-access coded caching (MACC) problem which consists of a central server with $N$ files, $K$ caches, each with memory $M$ units and $K$ users, each one connected to $L$ consecutive caches with a cyclic wrap-around; see Figure~\ref{fig:problemsetupmultiaccess} for an illustration. Caches are populated with content related to the files and each user then requests one out of the $N$ files, and all the users' requests have to be served by a broadcast message from the central server with the help of caches. The goal is to minimize the central servers' transmission rate by designing efficient placement and delivery policies.

The optimal transmission rates for both the CC and the MACC problems have been well-studied in the literature with both achievability and information-theoretic converses being derived; see for example  \cite{maddah2014fundamental,wan2016optimality,hachem2017codedmulti,reddy2020rate,sasi2020improved,serbetci2019multi,paschos2018role,cheng2021novel} for more details. Schemes with order-optimal rates usually require the files to be divided into many smaller subfiles (referred to as \textit{subpacketization}), with the number of subfiles often scaling exponentially with the system size. This in turn introduces large overhead which can eclipse the bandwidth gains that coded delivery schemes provide and hence there has been a lot of interest in designing low subpacketization schemes for both the CC and MACC problems \cite{chittoor2021subexponential,shanmugam2017coded,aravind2022lifting,tang2018coded,salehi2019subpacketization,cheng2017coded,sasi2021multi,wang2022multi} which incur as low a penalty in terms of the server transmission rate. 

\remove{
To achieve minimum transmission rate, \cite{maddah2014fundamental,wan2016optimality,hachem2017codedmulti,reddy2020rate,sasi2020improved,serbetci2019multi} divides each file into a lot of subfiles. If we split a file into $F$ subfiles, including both placement and delivery phases, then it is called $F$-subpacketization. In \cite{maddah2014fundamental,wan2016optimality,hachem2017codedmulti,reddy2020rate,sasi2020improved,serbetci2019multi}, $F$ grows exponentially in $K$. A high subpacketization requirement poses multiple issues like it requires large overhead size. Hence, recent works like \cite{chittoor2021subexponential,shanmugam2017coded,aravind2022lifting,sasi2021multi,wang2022multi} focus on the schemes with low sub-packetization. 
}
In this paper, we  focus on the MACC problem with linear subpacketization, i.e., the number of sub-files scales linearly with the number of users $K$. There has been some recent work on this problem. In particular, \cite{sasi2021multi} provide linear subpacketization schemes based on the idea of Placement Delivery Arrays (PDAs). However, these combinatorial structures only exist for a restricted set of system parameters. On the other hand, while \cite{wang2022multi} provide more general constructions, the subpacketization required is sometimes super-linear ($O(K^2)$). We provide alternate low subpacketization schemes for the MACC problem and are able to improve upon these prior results, as we discuss below. 

In particular, we use the same placement strategy as in \cite{wang2022multi} but propose a different delivery policy based on connections to the Index Coding Problem (ICP) \cite{birk1998informed,bar2011index} to get an improved transmission rate. If we fix the placement policy and the user requests, then solving a coded caching problem is equivalent to solving an index coding problem (ICP). This relation has proved instrumental in the literature both for designing better achievability schemes \cite{reddy2020rate, reddy2020structured} as well as proving converse results \cite{wan2016optimality}. In our policy, we map our MACC problem to a `structured' ICP studied previously in \cite{reddy2020structured} and are able to utilize the results there as well as new upper bounds derived here to establish new improved bounds on the optimal server transmission rate for the MACC problem under linear subpacketization. The main contributions of our paper are summarized below:
\begin{itemize}
	\item We derive new upper bounds on the optimal transmission rate for a structured ICP, which has been studied in the literature.
	\item We propose a policy with linear subpacketization for the MACC problem based on the mapping to this structured ICP, and are able to upper bound the transmission rate of our policy for any given system parameters.
	\item We compare our results with those in the literature and demonstrate improvement in terms of both the servers' transmission rate as well as the subpacketization level.
	\item Even though we mainly focus on MACC with linear subpacketization, we also provide improved results for $O(K^2)$  subpacketization using our policy.
\end{itemize}
The rest of the paper is organized as follows. Sections \ref{sec:not} and \ref{sec:setting} introduce the required notations and describe the formal problem setting. Section \ref{sec:ICP_preliminaries} discusses the ICP preliminaries. Section \ref{sec:policy} discusses our policy and Section \ref{sec:results} discusses the main results of our paper briefly. We conclude with a brief discussion in Section \ref{sec:conclusions}. All the proofs and an illustrating example are relegated to the Appendix.

%% file: setting.tex
\section{Notations}\label{sec:not}

\begin{itemize}
	\item $[m]=\{1,2,3,...,m\}$
	\item For some given{\footnote{In our manuscript, $K$ is the number of users in the index coding problem / multi-access coded caching problem.}}  integer $K$ such that $K\geq m, n$, \begin{align*}
		[n:m]=\begin{cases}
			\{n,n+1,...,m\} & \text{ if } n\leq m \\
			\{n,n+1,...,K,1,2,...,m\} & \text{ if } n>m 
		\end{cases}
	\end{align*}
	\item \begin{equation*}
		<n>_m=\begin{cases}
			n\Mod{m} & \text{ if } n\Mod{m}\neq 0 \\
			m & \text{ if } n\Mod{m}=0 
		\end{cases}
	\end{equation*}
	\item $|S|-$ size of file / subfile / set $S$ 
	\item {$\mathcal{F}_{j,\mathcal{S}}$} denotes parts of File $j$ exclusively available to users with index in set $\mathcal{S}$	
%	\item $len(\mathbf{u})$ denotes the number of components  of the vector $\mathbf{u}$
%	\item $\mathbf{\widehat{a}}$ denotes the maximum value amongst the components of the vector $\mathbf{a}$
\end{itemize}

\section{Setting} \label{sec:setting}
\begin{figure}[t]
	\begin{center}
		\includegraphics[scale=0.306]{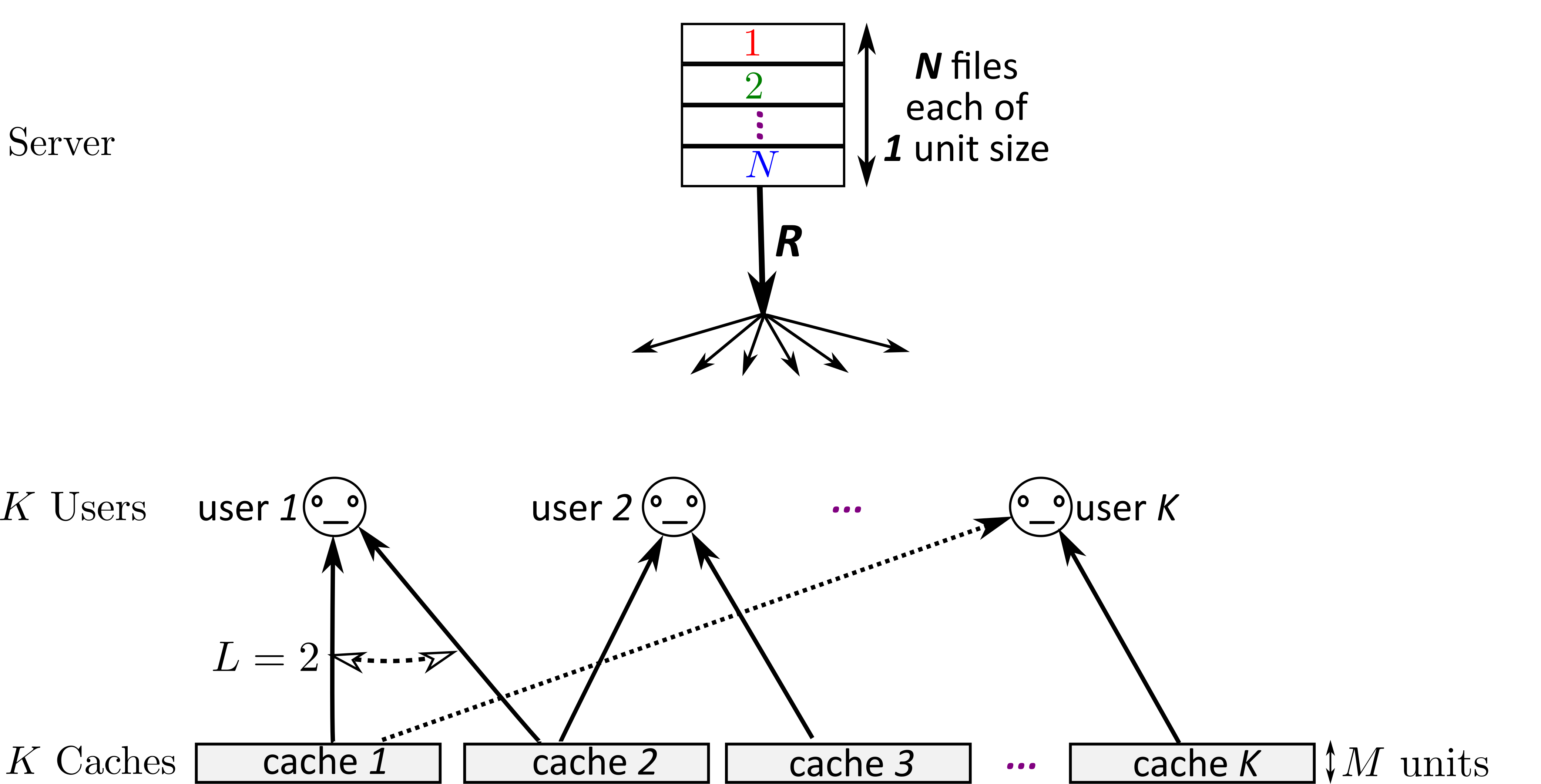}
		\caption{\sl An illustration of MACC consisting of a central server with $N$ files, $K$ caches, each of size $M$ units, and $K$ users, each user is connected to $L=2$ caches with a cyclic wrap around.   \label{fig:problemsetupmultiaccess}}
	\end{center}
	\setlength{\belowcaptionskip}{-1in}
\end{figure}

We focus on the MACC setup initially studied in \cite{hachem2017codedmulti}, which consists of a central server, $K$
caches, and $K$ users, see Figure \ref{fig:problemsetupmultiaccess}. We assume 
\begin{itemize}
	\item the central server has $N$ files $\mathcal{F}_1$, $\mathcal{F}_2, \ldots, \mathcal{F}_N,$ each of size $1$ unit ($=B$ bits{\footnote{We assume that $B$ is sufficiently large}})
		\item each cache has $M$ units of memory, 
	\item each user has access to $L$ consecutive caches with a cyclic wrap-around
	\item each user requests one out of the $N$ files, and
	\item the shared communication link between the central server and the users is error free.
\end{itemize}
%In this paper we consider a multi-access coded caching (MACC) problem with $N$ files and $K$ users. The files are denoted as $\mathcal{F}_1$, $\mathcal{F}_2, \ldots, \mathcal{F}_N.$ There are $K$ caches, each having a memory $M$, normalized to the file size. Each user accesses $L$ neighboring caches in cyclic wrap-around. 

Like the classical coded caching (CC) problem \cite{maddah-ali2013}, the MACC problem has two phases: a placement phase and a delivery phase, as described below.
\subsubsection{Placement Phase}
In the placement phase, each cache stores content  related to the $N$ files. In this paper, we restrict our analysis to uncoded placement policies with linear sub-packetization which implies that we are allowed to split the files into $O(K)$ parts and store the file parts, but coding across the file parts is not allowed. 
%In the linear sub-packetization, the number of file parts is in the order of $K$.

%The placement phase occurs before the users reveal their requests.

%In placement phase, we design a storage policy, which decides the content $\mathcal{Z}_i$ $(=\phi_i(W_1,W_2,...,W_n))$ of $n$ files to be stored in Cache $i$ for all $i$. In this paper, we restrict to uncoded storage policies i.e., we allowed to split the files into parts and store the file parts, but coding across the file parts is not allowed. The major advantage of uncoded storage is that it can handle asynchronous demands without increasing communication rate \cite{}. Placement phase occurs before users reveal their requests.

After the placement phase, users reveal their requested files. Let  user $j$  requests file $d_j$ and $(d_1,d_2,\ldots,d_K)$ represents the request profile. 
\subsubsection{Delivery Phase}
In the delivery phase, depending on the user request profile and the contents of the caches, the central server broadcasts a coded message. By using the broadcasted message and the  cache content  at the accessible caches, each user aims to reconstruct its requested file. %In the placement phase, the parts of files are stored in the $K$ caches subject to the memory constraint. After the demands are revealed by the users, the server transmits linear combination of parts of files in order to satisfy the demands in the delivery phase. 

 The goal is to design placement and coded delivery policies in such a way that the overall  server transmission rate  is minimized, where the transmission rate is defined as the total number of bits transmitted by the server, normalized by the file size $B$. Note that if a policy divides a file into $F$ parts, including both placement and delivery phases, then we say that it employs $F$-subpacketization. In our paper, we focus on policies with linear subpacketization, i.e., policies with $F$ in the order of $K$. In this work, our main aim is to design caching and delivery policies with linear subpacketization and as small a server transmission rate as possible. 
 
 %Let {\color{red}$R_F^*(M)$ denote the minimum transmission rate required in a system with cache memory $M$ using a policy with subpacketization $F$. In this work, our main aim is to design efficient policies with linear subpacketization and thus upper bound $R_L^*(M)$, which will denote the optimal rate amongst policies with linear subpacketization.}

\subsection{Previous Results}
There are several works on linear subpacketization policies for the original CC setup ($L = 1$) \cite{chittoor2021subexponential,shanmugam2017coded,aravind2022lifting,tang2018coded,salehi2019subpacketization}, and also many works that focus on the MACC problem with no subpacketization restrictions \cite{hachem2017codedmulti, reddy2020rate, reddy2020structured,sasi2020improved,serbetci2019multi}. However, to the best of our knowledge, only \cite{sasi2021multi} and \cite{wang2022multi} focus on both, i.e., MACC with linear sub-packetization and their main results are discussed below.
\remove{
many works on the Ali-Niesen coded caching setup ($L=1$) with linear sub-packetization, for example see \cite{chittoor2021subexponential, aravind2022lifting} and many works focus on MACC with no restrictions on subpacketization, for example see \cite{hachem2017codedmulti, reddy2020rate, reddy2020structured}. But, as per our knowledge only \cite{sasi2021multi} and \cite{wang2022multi} focus on both, i.e., multi-access coded caching with linear sub-packetization and their main results are discussed below:
}

The following lemma describes the main result of \cite{sasi2021multi}.
\begin{proposition}{\cite[Theorem 1]{sasi2021multi}}
	Consider a MACC with $N$ files, $K$ caches, each with memory  $M$ units and $K$ users, each user has access to $L$ consecutive caches with  a cyclic wrap-around. For $M\in \{iN/K:i\in [{\lfloor K/L \rfloor]}, i|K, (K-iL+i)|K\}$, we can achieve $R_1(M)$ using linear sub-packetization, where 
	\begin{align}
		R_{1}(M)=\frac{(K-iL)(K-iL+i)}{2K}.
	\end{align}\label{lem:BSR_UB}
\end{proposition}
The following lemma describes the main result of \cite{wang2022multi}.
	\begin{proposition}{\cite[Theorem 2]{wang2022multi}}
		Consider a MACC with $N$ files, $K$ caches, each with memory  $M$ units and $K$ users, each user has access to $L$ consecutive caches with  a cyclic wrap-around. For $M\in \{iN/K:i\in [{\lfloor K/L \rfloor]}\}$, we can achieve $R_2(M)$ using $F_2(M)-$sub-packetization, where 
	\begin{align}
		&R_2(M)= \nonumber\\
		&\begin{cases}
			\frac{(K-iL)(K-iL+1)}{2K} & \text{ if } (K-iL+1)|K \text{ or } K-iL=1 \\
			\frac{K-iL}{2\lfloor \frac{K}{K-iL+1} \rfloor + 1 } & \text{else if } <K>_{K-iL+1}=K-iL\\
			\frac{K-iL}{2\lfloor \frac{K}{K-iL+1} \rfloor  } & \text{ else}
		\end{cases}
	\end{align} and 
\begin{align}
	&F_2(M)= \nonumber\\
	&\begin{cases}
		{K} & \text{ if } (K-iL+1)|K \text{ or } K-iL=1 \\
		\big({2\lfloor \frac{K}{K-iL+1} \rfloor + 1 }\big)K & \text{else if } <K>_{K-iL+1}=K-iL\\
		\big({2\lfloor \frac{K}{K-iL+1} \rfloor  }\big)K & \text{ else}.
	\end{cases}
\end{align}\label{lem:chinese_ub}
	\end{proposition}
%
%Note that, Proposition \ref{lem:chinese_ub} tells that the sub-packetization in \cite{wang2022multi} is sometimes in the order of $K^2$.
%
Note that the results in \cite{sasi2021multi} are applicable to a restricted set of system parameters. On the other hand, while \cite{wang2022multi} provide more general results, the subpacketization required is sometimes super-linear ($O(K^2)$). In our paper, we focus on multi-access coded caching with linear subpacketization for all system parameters. Moreover, in certain parameter regions, our results perform better than the earlier results. The following example illustrates this fact.
\begin{example}
	Consider a MACC problem with $N=K=8$, $L=2$ and $M=i=3$. Proposition \ref{lem:BSR_UB} doesn't apply to these parameters and Proposition \ref{lem:chinese_ub} achieves $R_2(M)=0.5 $ units using subpacketization $F=4K=32$. But, our policy can achieve a rate of 3/8 units using subpacketization $F=K=8$, see Example \ref{ex:detailed_example}  for details.
\end{example}

%% file: preliminaries.tex
\section{Index Coding Problem -- Preliminaries}\label{sec:ICP_preliminaries}
In a coded caching problem, if we fix the placement policy and the user request profile, then solving the coded caching problem is equivalent to solving an ICP \cite{bar2011index,birk1998informed}, which is well studied in the literature. Our proposed policy exploits this connection and we design delivery policies based on the solution of an ICP. Hence, we first see some preliminaries regrading ICP and existing results which prove useful for our work. In particular, we briefly focus on a structured ICP which was studied in  \cite{reddy2020structured} for deriving improved rates for the MACC problem. In our work, we also derive  better results on this structured ICP to improve the state of the art; these results are discussed at the end of the section.

An ICP consists of a server possessing $N$ files and $K$ users demanding some subsets of the files. Each user possesses a subset of files as side-information. Knowing the users' side-information and demand, the server broadcasts linear combinations of files with which the users can decode their demanded files. The goal is to minimize the servers transmission rate.  Let the set of files be $\mathcal{F}=\{x_1$, $x_2, \ldots, x_N\} $ and the set of $K$ users be $\mathcal{U}=\{U_1, U_2, \ldots, U_K\}$. Each user is represented as $U_k=(\mathcal{K}_k, \mathcal{W}_k)$, where $\mathcal{K}_k \subset \mathcal{F}$ is the known-set or the side-information set and $\mathcal{W}_k \subseteq \mathcal{F}$ is the want-set or the demand set of the $k$-th user. The interference set of $U_k$ is $\mathcal{I}_k=(\mathcal{K}_k \cup \mathcal{W}_k)^c.$ Even though finding the optimal transmission rate of a general ICP is still open problem, there are several upper and lower bounds available in the literature. Some of the upper bounds are based on min-rank \cite{bar2011index}, local chromatic number, fractional local chromatic number \cite{shanmugam2013local}, and local partial clique cover \cite{agarwal2016local}. Some of the lower bounds are based on  maximum induced acyclic sub-graph (MAIS) \cite{arbabjolfaei2013capacity}, conflict distance \cite{jafar2014interference}, and dimension counting \cite{maleki2014index}.

A unicast index coding problem (UICP) is an ICP in which the same file is not demanded by more than one user. A single unicast index coding problem (SUICP) is  a UICP in which each user demands only a single file. We refer to an SUICP with $K$ users as {an $(a_1,a_2)_z-$ICP} for some  non-negative integers $a_1,a_2$ and a natural number $z$ such that 
\begin{align}\label{eq:sumcondn}
	a_1+a_2+z=K-1,
\end{align} if $\forall k\in[K]$, User $k$
\begin{itemize}
	\item  wants $x_k$, i.e., $\mathcal{W}_k=\{x_k\}$ and
	\item knows 
		$\mathcal{K}_k=\{x_b:b=<k+a_1+r>_K, r\in[z]\}$, \\
	i.e., the side-information $\mathcal{K}_k$ is a collection of $z$ consecutive elements. More explicitly, the side information of User $k$ is given by
	$$\mathcal{K}_k=\{x_{<k+a_1+1>_K},x_{<k+a_1+2>_K},...,x_{<k+a_1+z>_K}\}.$$
\end{itemize}
The following result provides an upper bound on the transmission rate of an $(a_1,a_2)_z-$ICP. 
\begin{proposition}\cite[Proposition 4]{reddy2021optimal}
	Let $R^*$ be the optimal transmission rate of an $(a_1,a_2)_z-$ICP, then $$R^*\leq R =a_1+a_2+1.$$
\end{proposition}

 Let there be $K$ users such that User $k$ wants $\mathcal{W}_k^1$ and knows $\mathcal{K}_k^1$ in the first instance and the same User $k$ wants $\mathcal{W}_k^2$ and knows $\mathcal{K}_k^2$ in the second instance. By assuming no common files involved across the two instances,  we define the union of these two instances is an ICP with 
\begin{itemize}
	\item want set $\mathcal{W}_k=\mathcal{W}^1_k\cup \mathcal{W}^2_k$ and
	\item known set $\mathcal{K}_k=\mathcal{K}^1_k\cup \mathcal{K}^2_k$.
\end{itemize}
We can generalize this to union of more instances also.

For some  non-negative integers $a_1,a_2(\leq a_1)$ and a natural number $z$, the union of $(a_1,a_2)_z-$ICP and  $(a_2,a_1)_z-$ICP is referred as the $\overline{(a_1,a_2)}_z-$ICP. Let $x_{k,j}$ be the $j^{\text{th}}$ file requested by User $k$. In the $\overline{(a_1,a_2)}_z-$ICP, $\forall k\in[K]$, User $k$
\begin{itemize}
	\item want set $\mathcal{W}_k=\{x_{k,1},x_{k,2}\}$
	\item known set $$\mathcal{K}_k=\{x_{b,t}:b= <k+a_t+r>_K, t\in[2], r\in[z]\}.$$
\end{itemize}
 See \cite{reddy2020structured} for more details. Note that $(a_1,a_2)_z-$ICP is an $(a_1,a_2,...,a_i)_{z}-$ICP and $\overline{(a_1,a_2)}_z-$ICP is an $\overline{(a_1,a_2,...,a_i)}_{z}-$ICP with $i=2$, which are studied in \cite{reddy2020structured}.
The following lemma from \cite{reddy2020structured} gives an upper bound as well as a lower bound on the number of optimal transmission rate of $\overline{(a_1,a_2)}_z-$ICP.
\begin{proposition}\cite[Theorem 1, Theorem 2 and Theorem 3]{reddy2020structured}
	 For some  non-negative integers $a_1,a_2 (\leq a_1)$ and a natural number $z$, consider an $\overline{(a_1,a_2)}_z-$ICP with $K$ users. Let $\overline{R}^*$ be the optimal transmission rate of the ICP, then $$\overline{R}^*\leq \overline{R}_1=\begin{cases}
	 	a_1+a_2+2 & \text{ if } (a_1+a_2+2)|K \\
	 	\min\{a_1+2a_2+2,K\} & \text { otherwise}
	 \end{cases}$$
  and $$\overline{R}^*\geq \overline{R}_l= a_1+a_2+2.$$\label{lem:a1a2ICP_UB1} 
\end{proposition}
In the following theorem, we give a new upper bound on $\overline{(a_1,a_2)}_z-$ICP  transmission rate.
\begin{theorem}\label{thm:newlocalchromatic}
	Let $\overline{R}^*$ be the optimal transmission rate of $\overline{(a_1,a_2)}_z-$ICP, for some  non-negative integers $a_1,a_2 (\leq a_1)$ and a natural number $z$. If there exists an integer $\overline{R}_2$ such that $\overline{R}_2 \geq a_1+a_2+2$ and $\overline{R}_2 | K$, then $\overline{R}^*\leq \overline{R}_2$.
\end{theorem}

The following example compares the results of Proposition \ref{lem:a1a2ICP_UB1} and Theorem \ref{thm:newlocalchromatic}.

\begin{example}
	For $\overline{(2,2)}_{9}-$ICP, we have $K=14$ users and the lower bound $\overline{R}_l=6$ units and the upper bounds $\overline{R}_1=8$ units, $\overline{R}_2 =7$ units. We can clearly see that $\overline{R}_2 < \overline{R}_1$. 
\end{example}

The upper bounds in Proposition \ref{lem:a1a2ICP_UB1} and Theorem \ref{thm:newlocalchromatic} are based on the local chromatic number \cite[Theorem 1]{shanmugam2013local} and correspond to scalar index coding.  We can improve the achievable rate by using vector index coding via the fractional local chromatic number \cite[Theorem 2]{shanmugam2013local} based upper bound;  the improved bound is stated below:
\begin{theorem}
	For some  non-negative integers $a_1,a_2(\leq a_1)$ and a natural number $z$, consider an $\overline{(a_1,a_2)}_z-$ICP with $K$ users. Let $\overline{R}^*$ be the optimal transmission rate of the ICP, then $$\overline{R}^*\leq \overline{R}_3=\frac{\min\{\lfloor \frac{K}{a_1+a_2+2} \rfloor (a_1+a_2+2)+a_2, K  \} }{\lfloor \frac{K}{a_1+a_2+2} \rfloor}.
	$$\label{thm:a1a2ICP_UB2} 
\end{theorem}
The following example compares Proposition \ref{lem:a1a2ICP_UB1} and Theorem \ref{thm:a1a2ICP_UB2}.
\begin{example}
	For $\overline{(10,10)}_{1000}-$ICP, we have $K=1021$ users and the lower bound $\overline{R}_l=22$ units and the upper bounds $\overline{R}_1=32$ units, $\overline{R}_3 \approx 22.2173$ units. We can clearly see that $\overline{R}_3 < \overline{R}_1$ and $\overline{R}_3$ is approximately equal to $\overline{R}_l$.
\end{example}

%% file: Ourpolicy.tex
\section{Our Policy}\label{sec:policy}
Now, we discuss our policy for the MACC problem with linear subpacketization. Our policy is proposed for corner points $M=iN/K$, where {$i\in [\lceil K/L\rceil ] \cup \{0\}$}. For the intermediate values, we use memory sharing appropriately. As we mentioned earlier, there are two phases in our policy. The first phase is the placement phase and the second phase is delivery phase. In our paper, we use the same placement policy as in \cite{wang2022multi}, but use a different delivery policy to improve the previous results. Our delivery policy uses the upper bounds on $\overline{(a_1,a_2)}_z-$ICPs mentioned in Section \ref{sec:ICP_preliminaries}.

\subsection{Placement Phase:}  We can achieve the transmission rates $K$ and $0$ at the trivial corner points $M=0$ and {$\lceil K/L\rceil $} respectively.  Now, we briefly describe our placement policy  for the non-trivial memory points  $M=iN/K$, where $i\in [\lfloor K/L\rfloor ]$. 

We divide each file $\mathcal{F}_n$ $n\in [N]$ into $K$ disjoint subfiles $\mathcal{F}_{n,m}$ $m\in[K]$  of equal size. subfile $\mathcal{F}_{n,m}$ is stored in caches $m, m+L, ..., m+(i-1)L$. Hence, cache $k$, $k\in[K]$ stores the  subfiles $\mathcal{M}_k=\{\mathcal{F}_{n,<k-(r-1)L>_K}:n\in[N],r\in[i]\}$. 

Some useful properties of the placement policy are
\begin{itemize}
	\item each cache stores $i$ subfiles of each file. Hence, the memory constraint $M=iN/K$ is satisfied. 
	\item no two caches store the same set of subfiles
	\item each subfile is available to exactly $iL$ consecutive users and no two subfiles are available to the  same set of caches
	\item each user has access to $iL$ subfiles and no two users have the  same set of available subfiles
\end{itemize} 

Since, subfile $\mathcal{F}_{n,m}$ is stored in caches $m, m+L, ..., m+(i-1)L$, it is available to users with the indices in the set $[<m-L+1>_K:<m+(i-1)L>_K]$. Hence, $\mathcal{F}_{n,m}$ is represented as $\mathcal{F}_{n,[<m-L+1>_K:<m+(i-1)L>_K]}$, where the indices in the subscripted set indicates that the subfile available at corresponding users. For each subset of the set $[K]$ with $iL$ circular consecutive elements, there exists a subfile and vice versa. Note that there are a total of $K$ subsets and $K$ subfiles.

\subsection{Delivery Phase:} In this phase, each user reveals their request. Let the request pattern be $\mathbf{d}=(d_1,d_2,...,d_K)\in[N]^K$, i.e., user $j$ requests file $\mathcal{F}_{d_j}$. In other words, user $j$  requests $K$ subfiles of file $\mathcal{F}_{d_j}$. Due to our placement policy, user $j$ already has access to $iL$ subfiles and it only needs remaining $K-iL$ subfiles of file $\mathcal{F}_{d_j}$.
In particular, user $j$ is connected to caches in the set $[j:j+L-1]$ and cache $k$ for $k\in[K]$ stores the  subfiles $\mathcal{M}_k=\{\mathcal{F}_{n,<k-(r-1)L>_K}:n\in[N],r\in[i]\}$. Hence, user $j$ already has access to $\mathcal{F}_{d_j,<j+l-1-(r-1)L>_K}$, $r\in[i]$,  $l\in[L]$ and needs the remaining $K-iL$ subfiles. 

Now, we form an instance of ICP with $K$ users, such that each user $j$ wants the unavailable $K-iL$ subfiles of file $\mathcal{F}_{d_j}$. The side-information of user $j$ is the subfiles needed for the remaining $K-1$ users and stored in user $j$'s available caches $[j:j+L-1]$. The central server transmits messages according to the solution of the ICP. Note that we show that the ICP formed is the union of several $\overline{(a_1,a_2)}_{z=iL}-$ICPs and use the results in Section \ref{sec:ICP_preliminaries} to get an upper bound on the optimal transmission rate of the ICPs. Finally, the optimal transmission rate of our MACC ICP is upper bounded by the sum of the upper bounds for the individual $\overline{(a_1,a_2)}_{z=iL}-$ICPs, see Appendix \ref{sec:proof} for details.

%% file: mainresults.tex
\section{Main Results}\label{sec:results}
In this section, we characterize the transmission rate of our policy. First, we discuss an improved upper bound on the MACC transmission rate with linear subpacketization, which is a generalization of Proposition \ref{lem:BSR_UB}.
\begin{theorem}
	Consider the MACC with $N$ files, $K$ caches, each with memory  $M$ units and $K$ users, each user has access to $L$ consecutive caches with  a cyclic wrap-around. For $M\in \{iN/K:i\in [0,K/L]\}$, if there exists $X\in \mathbb{N}$ such that $X\geq K-iL+1$ and $X|K$, then we can achieve $R_3(M)$ using linear sub-packetization, where 
	\begin{align}
		R_3(M)=\frac{(K-iL)X}{2K}.
	\end{align}\label{thm:BSR_UB_gen}
\end{theorem}
If $i$ and $K-iL+i$ divide $K$, by substituting $X=K-iL+i$, we recover Proposition \ref{lem:BSR_UB} from Theorem \ref{thm:BSR_UB_gen}. To achieve these results,  we use $F=K$ if $K-iL$ is even and $F=K+1$ if $K-iL$ is odd; both are linear in $K$.

Now, we discuss  another upper bound on the transmission rate for the general MACC problem with linear subpacketization and no restrictions on the parameters.  Let {$R_{\text{s}}(M)$, $R_{\text{e}}(M)$ and $R_{\text{o}}(M)$}, at $M=iN/K$, $i\in [\lfloor K/L\rfloor ]$  be defined as follows:
\begin{equation}\label{eq:tight}
	{R_{\text{s}}(M)}:=\frac{(K-iL)(K-iL+1)}{2K}
\end{equation}

\begin{equation}\label{eq:even_upperbound}
	{R_{\text{e}}(M)}:=\begin{cases}
	\frac{K(K-iL)-iL(iL-1)}{2K}	 & \text{ if } K\geq 3iL\\
	\frac{(K-iL)(5K-5iL+2)}{8K}	 & \text{ otherwise } 
	\end{cases}
\end{equation}
\begin{equation}\label{eq:odd_upperbound}
	{R_{\text{o}}(M)}:=\begin{cases}
		\frac{K(K-iL+1)-iL(iL+1)}{2K}	 & \text{ if } K\geq 3iL+1\\
		\frac{(K-iL-1)(5K-5iL+9)}{8K}+1 & \text{ otherwise } 
	\end{cases}
\end{equation}
The following theorem discusses an upper bound on the transmission rate for the general multi-access coded caching with linear subpacketization.
	\begin{theorem}\label{thm:our_linear_results}
	Consider a MACC with $N$ files, $K$ caches, each with memory  $M$ units and $K$ users, each user has access to $L$ consecutive caches with  a cyclic wrap-around. For $M\in \{iN/K:i\in [0,\lfloor K/L \rfloor]\}$, let $R_\text{s}(M)$, $R_{\text{e}}(M)$ and $R_{\text{o}}(M)$ be defined as in \eqref{eq:tight}, \eqref{eq:even_upperbound} and \eqref{eq:odd_upperbound} respectively. Then, we can achieve {$R_4(M)$} using linear sub-packetization, where
	\begin{align*}
		{R_4(M)}= 
		\begin{cases}
			R_\text{s}(M)	 & \text{ if } (K-iL+1)|K \text{ or } K-iL=1  \\
			R_{\text{e}}(M) & \text{else if } K-iL \text{ is even }\\
			R_{\text{o}}(M) & \text{ else}
		\end{cases}
	\end{align*} 
\end{theorem}
{
If we consider the case where $iL<K/3$, $<K>_{K-iL+1}\neq K-iL$, and $K-iL$ is even, then from Proposition \ref{lem:chinese_ub}, we have $R_2(M)=\frac{K-iL}{2}$ while from Theorem \ref{thm:our_linear_results}, we have $R_4(M)=R_{\text{e}}(M)=\frac{K-iL}{2}-\frac{iL(iL-1)}{2K}<R_2(M)$. This shows that Theorem \ref{thm:our_linear_results} results can be better than Proposition \ref{lem:chinese_ub} results. Moreover, Proposition \ref{lem:chinese_ub} uses non-linear subpacketization for some parameters, whereas Theorem \ref{thm:our_nonlinear_results} uses only linear subpacketization for all the parameters.  
}

If we increase the allowed subpacketization from linear to $O(K^2)$, we can further improve the results of Theorem \ref{thm:our_linear_results} without changing the placement policy. Let $\widetilde{R}_{\text{e}}(M)$ and $\widetilde{R}_{\text{o}}(M)$, at $M=iN/K$, $i\in [\lfloor K/L\rfloor ]$  be defined as follows:

\begin{align}\label{eq:even_improvedbound1}
	\widetilde{R}_{\text{e}}(M):= 
	\sum_{j=0}^{\frac{K-iL}{2}-1}	\frac{\min\{\lfloor \frac{K}{K-iL+1} \rfloor (K-iL+1)+j, K  \}}{\lfloor \frac{K}{K-iL+1} \rfloor K}	\
\end{align}

\begin{align}\label{eq:odd_improvedbound1}
	\widetilde{R}_{\text{o}}&(M):= \frac{\min\{\lfloor \frac{K}{K-iL+1} \rfloor (K-iL+1)+\frac{K-iL-1}{2}, K  \}}{2 \lfloor \frac{K}{K-iL+1} \rfloor K}	\nonumber \\
	&+	\sum_{j=0}^{\frac{K-iL-3}{2}}	\frac{\min\{\lfloor \frac{K}{K-iL+1} \rfloor (K-iL+1)+j, K  \}}{\lfloor \frac{K}{K-iL+1} \rfloor K}	
\end{align} 
The improved results with $O(K^2)$-subpacketization are stated below:
	\begin{theorem}\label{thm:our_nonlinear_results}
	Consider a MACC with $N$ files, $K$ caches, each with memory  $M$ units and $K$ users, each user has access to $L$ consecutive caches with  a cyclic wrap-around. For $M\in \{iN/K:i\in [0,\lfloor K/L \rfloor]\}$, let  $\widetilde{R}_{\text{e}}(M)$ and $\widetilde{R}_{\text{o}}(M)$ be defined as in  \eqref{eq:even_improvedbound1} and \eqref{eq:odd_improvedbound1} respectively. Then, we can achieve $\widetilde{R}_5(M)$  using $\widetilde{F}_5(M)-$sub-packetization, where 
	\begin{align}
		\widetilde{R}_5(M)= 
		\begin{cases}
			\widetilde{R}_{\text{e}}(M) & \text{ if } K-iL \text{ is even }\\
			\widetilde{R}_{\text{o}}(M) & \text{ otherwise}
		\end{cases}
	\end{align} and 
	\begin{align}
		&\widetilde{F}_5(M)= \nonumber\\
		&\begin{cases}
			\widetilde{K} & \text{ if } (K-iL+1)|K \text{ or } K-iL =1 \\
			{\widetilde{K}\lfloor \frac{K}{K-iL+1} \rfloor  } & \text{ else}
		\end{cases}
	\end{align}
where,
\begin{align*}
	\widetilde{K}= 
	\begin{cases}
		K & \text{ if } K-iL \text{ is even or } K-iL =1 \\
		K+1 & \text{ else}.
	\end{cases}
\end{align*}
\end{theorem}

Below, we present an example to compare our results with those available in the literature. 

\begin{example}
	Consider a MACC with $K=100$, $i=20$, $L=4$, we have the transmission rates $R_2(M)=R_3(M)=2.5$, {${R}_4(M)= 2.55$,} and $\widetilde{R}_5(M)\approx 2.14$ and the subpacketization $F_2(M)=800 (>K)$ and $\widetilde{F}_5(M)=400 = F_2(M)/2$. Note that $R_1(M)$ doesn't exist for above parameters.
\end{example}
 This example shows that our results can improve upon those in earlier works, both in terms of transmission rate and subpacketization.
 
 \begin{remark}
 	By substituting $L=1$, we can use our results for the original CC problem as well to derive improved rates with linear subpacketization. %We can also show that for our placement policy, our results are orderwise optimal. %{\color{red}For Ali-Niesen coded caching setup also, these are the first results on linear subpacketization with no restrictions on parameters. }
 \end{remark}
 
 \begin{remark}
 	As mentioned before, we use bounds on the local chromatic number to bound the transmission rate of the structured ICPs that we map the MACC problem to.  One potential issue with this is that we need the field size to be large to construct MDS codes, see \cite{shanmugam2013local} for details. This may not be always be feasible and a natural question is whether we can construct binary codes which provide comparable performance. One partial resolution is provided by \cite[Corollary 1]{shanmugam2013local}, which implies that with high probability, we can construct binary MDS codes for our MACC problem by adding an extra transmission rate of {$\frac{2\log (K(K-iL))}{K}$} units. 
 \end{remark}

\section{Conclusions}\label{sec:conclusions}
In this paper, we focus on the design and analysis of caching and delivery schemes for the MACC problem with linear subpacketization. These schemes and the corresponding upper bounds on the server transmission rate are based on the connection of the coded caching problem with a previously studied structured index coding problem, for which we derive  new improved bounds that are of interest in their own right. Our results improve upon the state of the art for several parameter regimes. Deriving the lower bound on the transmission rate and characterizing the exact rate-memory trade-off for MACC with linear subpacketization is part of our future work. 

%% file: proofs.tex
\section{Proof of Theorem \ref{thm:newlocalchromatic}}
We use the same proof technique of \cite[Thoerem 1]{reddy2020structured}. % 
We use the following definitions of ICP terms from \cite[Section IV]{reddy2020structured} for our proof.  If user $j$'s requested file is not available at user $k$, then user $j$ is an interfering user to user $k$. The set containing user $k$ and all its interfering users are called  the closed anti-outneighborhood of user $k$. A coloring scheme assigns a color to each user and is said to be 'proper coloring scheme', if it assigns colors such that no user shares its color with any of its interfering users. The local chromatic number $\mathcal{X}_l$ of an ICP is defined as the maximum number of different colors that appear in the closed anti-outneighborhood of any user, minimized over all proper coloring schemes. We also use the following lemma in our proof.
\begin{lemma} \cite[Theorem 1]{shanmugam2013local} Let $\mathcal{X}_l$ be the local chromatic number of the side information graph of an ICP, then  the optimal transmission rate of the ICP is upper bounded by $\mathcal{X}_l$. \label{lem:lovalchromatic_ub}
\end{lemma}
Now, we discuss the proof of  Theorem \ref{thm:newlocalchromatic}.
\begin{proof}
Recall from Section \ref{sec:ICP_preliminaries} that in the $\overline{(a_1,a_2)}_z-$ICP, $\forall k\in[K]$, User $k$
\begin{itemize}
	\item want set $\mathcal{W}_k=\{x_{k,1},x_{k,2}\}$
	\item known set $$\mathcal{K}_k=\{x_{b,t}:b= <k+a_t+r>_K, t\in[2], r\in[z]\}.$$
\end{itemize}
i.e., in an $\overline{(a_1,a_2)}_z-$ICP, each user wants $2$ files and has $2z$ files as side-information. We form a $K\times 2$ table such that $p^{\text{th}}$ row and $q^{\text{th}}$ column contains $q^{\text{th}}$ file requested by User $p$, see Table \ref{Tab:genicp}. We denote the $p^{\text{th}}$ row and $q^{\text{th}}$ column's entry as Node $(p,q)$.

\begin{table}[h]
	%\begin{minipage}{.43\linewidth}
	\vspace{0.8cm}
	%\caption{}
	\centering
	%\resizebox{\linewidth}{!}{
		\begin{tabular}{| c | c | }
			\hline 
			$x_{1,1}$ &  $x_{1,2}$\\
			\hline 
			$x_{2,1}$ &  $x_{2,2}$ \\
			\hline 
			\vdots & \vdots  \\
			\hline
			$x_{K,1}$ &  $x_{K,2}$ \\
			\hline 
					\hline
					$(a_1,a_2)_z-$ICP & $(a_2,a_1)_z-$ICP \\
					\hline
		\end{tabular}
		\vspace{0.2cm}
		\caption{\sl The tabular representation of the  $\overline{(a_1,a_2)}_L-$ICP. $x_{p,q}$ represents the $q^{\text{th}}$ file requested by User $p$. In the table, the first column corresponds to the $(a_1,a_2)_z-$ICP and the second column corresponds to the $(a_2,a_1)_z-$ICP.  } \label{Tab:genicp}
	\end{table}

	Recall that the $\overline{(a_1,a_2)}_z-$ICP is a UICP with $K$ users, and each user requesting 2 distinct files. We convert this UICP into an SUICP with $2K$ virtual users, each one requesting  a distinct file. In particular, each user in the UICP is  mapped to 2 virtual users in the SUICP, such that each virtual user requesting a distinct file of the original user's requested files.  The side-information at the virtual user is the same as 
	its corresponding original user. Note that each node represents a virtual user in the constructed SUICP. In Table \ref{Tab:genicp},  the side-information for Node $(p,q)$ in the constructed SUICP is the same as the side-information for User $p$ in the original UICP.  %is same as side-information of User $k$.
	 The optimal transmission rates for the both  UICP and constructed SUICP are same. Hence,
	we give an upper bound for the constructed SUICP, and it also works as a bound for the original UICP.

 \textit{A proper coloring scheme:}	We take $\bar{R}_2$ colors, and for Node ($k,1$), assign Color $<k>_{\bar{R}_2}$ and for Node ($k,2$), assign Color $<k+a_1+1>_{\bar{R}_2}$, $\forall k\in[K]$. Note that every color occurs in a column exactly $K/\bar{R}_2$ times. Now, we verify that it is a proper coloring scheme. 
 
 Consider a node from Column 1, say Node ($k,1$). Note that its color is $<k>_{\bar{R}_2}$ and its interfering nodes in Column 1 are Nodes $(<k-j>_{K},1)$ $\forall j\in[a_2]$ and Nodes $(<k+j>_{K},1)$ $\forall j\in[a_1]$ and their colors are $<k-j>_{\bar{R}_2}$ $\forall j\in[a_2]$ and  $<k+j>_{\bar{R}_2}$ $\forall j\in[a_1]$ respectively. We can easily verify that none of these colors overlap with Color $<k>_{\bar{R}_2}$. Hence, we can say that Node ($k,1$) doesn't share its color with its  first column interference nodes. Node ($k,1$)'s interfering nodes in Column 2 are Nodes $(<k-j>_{K},2)$ $\forall j\in[a_1]$, Node ($k,2$), and Nodes $(<k+j>_{K},2)$ $\forall j\in[a_2]$ and their colors are $<k-j+a_1+1>_{\bar{R}_2}$ $\forall j\in[a_1]$, $<k+a_1+1>_{\bar{R}_2}$ and  $<k+j+a_1+1>_{\bar{R}_2}$ $\forall j\in[a_2]$ respectively. Again, we can easily verify that none of these colors are Color $<k>_{\bar{R}_2}$. Hence, we can say that Node ($k,1$) doesn't share its color with its  second column interference nodes either. Hence, we can say that under the proposed coloring scheme, none of the nodes of Column 1 share their colors with any of the corresponding interfering nodes. By using similar arguments, we can show that the same holds true for all nodes in Column 2 as well. Therefore, the coloring scheme is proper.

 Since, we are using only  $\bar{R}_2$ colors, every node contains at most $\bar{R}_2$ colors in its closed anti-outneighborhood. Therefore, from Lemma  \ref{lem:lovalchromatic_ub}, we have $R^*\leq \bar{R}_2$. 
 
\end{proof}
 
 \section{Proof of Theorem \ref{thm:a1a2ICP_UB2}}
 \begin{proof}
  If we split a file in the $(a_1,a_2)_z-$ICP into $m$ equal sized subfiles, then with
  respect to the subfiles, we have the union of $m$ $(a_1,a_2)_z-$ICPs, see \cite[Theorem 4 proof]{reddy2020structured} for details. To prove Theorem \ref{thm:a1a2ICP_UB2}, we divide the files in  the $\overline{(a_1,a_2)}_z-$ICP into $m=\lfloor\frac{K}{a_1+a_2+1}\rfloor$ subfiles. With respect to subfiles, it represents union of $m$ $\overline{(a_1,a_2)}_z-$ICPs and the following table represents it. %Its tabular representation is given in Table \ref{Tab:genicp_split}.
  
  \begin{table}[h]
  	%\begin{minipage}{.43\linewidth}
  	\vspace{0.8cm}
  	%\caption{}
  	\centering
  	%\resizebox{\linewidth}{!}{
  		\begin{tabular}{| c | c | c | c | c | c | c |}
  			\hline 
  			$x_{1,1}^1$ &  $x_{1,2}^1$ & $x_{1,1}^2$ &  $x_{1,2}^2$ & ... & $x_{1,1}^p$ &  $x_{1,2}^p$\\
  			\hline 
  			$x_{2,1}^1$ &  $x_{2,2}^1$ & $x_{2,1}^2$ &  $x_{2,2}^2$ & ... & $x_{2,1}^p$ &  $x_{2,2}^p$\\
  			\hline 
  			\vdots & \vdots & \vdots & \vdots & \vdots& \vdots & \vdots \\
  			\hline
  			$x_{K,1}^1$ &  $x_{K,2}^1$ & $x_{K,1}^2$ &  $x_{K,2}^2$ & ... & $x_{K,1}^p$ &  $x_{K,2}^p$ \\
  			\hline 
  		\end{tabular}
  		\vspace{0.2cm}
  		\caption{\sl The tabular representation of the  union of $m$ $\overline{(a_1,a_2)}_z-$ICPs. $x_{k,j}^m$ represents the $m^{\text{th}}$ subfile of $j^{\text{th}}$ file requested by User $k$. In the table, odd columns correspond to the $(a_1,a_2)_z-$ICP and even columns correspond to the $(a_2,a_1)_z-$ICP.  } \label{Tab:genicp_split}
  	\end{table}	
  
  Note that Table \ref{Tab:genicp_split} represents a UICP with $K$ users, and each user requesting $2m$ distinct subfiles. We convert this UICP into an SUICP with $2Km$ virtual users, each one requesting  a distinct subfile. In particular, each user in the UICP is  mapped to $2m$ virtual users in the SUICP, such that each virtual user requesting a distinct subfile of the original user's requested files.  The side-information at the virtual user is the same as 
  its corresponding original user. Note that in Table \ref{Tab:genicp_split}, each node represents a virtual user.
  We give an upper bound for the SUICP, and it also works as a bound for the original UICP in Table \ref{Tab:genicp_split}.
  
  \textit{A proper coloring scheme:}	We take $K$ colors, and for Node ($k,p$), assign Color $<k+(\frac{p-1}{2})(a_1+a_2+2)>_{K}$, if $p$ is odd and Color $<k-(\frac{p}{2}-1)(a_1+a_2+2)+a_2+1>_{K}$, if $p$ is even, $\forall k\in[K], p\in [2m]$. %and for Node ($k,2$), assign Color $<k+a_1+1>_{\bar{R}_2}$, $\forall k\in[K]$. Note that every color occurs in a column exactly $K/\bar{R}_2$ times. 
  Now, we verify that it is a proper coloring scheme. 
  
  Consider a node from Column 1, say Node ($k,1$). Note that its color is $k$ and its interfering nodes in odd column $p\in[2m]$ are Nodes $(<k-j>_{K},p)$ $\forall j\in[a_2]$, Node $(k,p)$ if $ p\neq 1$  and Nodes $(<k+j>_{K},p)$ $\forall j\in[a_1]$ and their colors are  $<k-j+(\frac{p-1}{2})(a_1+a_2+2)>_{K}$ $\forall j\in[a_2]$, $<k+(\frac{p-1}{2})(a_1+a_2+2)>_{K}$ if $p\neq 1$,  and  $<k+j+(\frac{p-1}{2})(a_1+a_2+2)>_{K}$ $\forall j\in[a_1]$ respectively. If $m=\lfloor\frac{K}{a_1+a_2+1}\rfloor$, we can easily verify that none of these colors overlap with Color $k$. Hence, we can say that Node ($k,1$) doesn't share its color with its  odd column interference nodes. Node ($k,1$)'s interfering nodes in even column $p\in[2m]$ are Nodes $(<k-j>_{K},2)$ $\forall j\in[a_1]$, Node ($k,2$), and Nodes $(<k+j>_{K},2)$ $\forall j\in[a_2]$ and their colors are $<k-j+(\frac{p}{2}-1)(a_1+a_2+2)+a_2+1>_{K}$ $\forall j\in[a_1]$, $<k+(\frac{p}{2}-1)(a_1+a_2+2)+a_2+1>_{K}$ and  $<k+j+(\frac{p}{2}-1)(a_1+a_2+2)+a_2+1>_{K}$ $\forall j\in[a_2]$ respectively. Again, we can easily verify that none of these colors are Color $k$. Hence, we can say that Node ($k,1$) doesn't share its color with its  even columns interference nodes either. Hence, we can say that under the proposed coloring scheme, none of the nodes of Column 1 share their colors with any of the corresponding interfering nodes. By using similar arguments, we can show that the same holds true for all nodes in other columns as well. Therefore, the coloring scheme is proper.
  
  According to the definition of $(a_1,a_2)_z-$ICP, if we arrange the users  in a $(m,n)_z-$ICP on a circle, a user is followed by its $m$ interference users then $z$ side-information  users and then it ends with $n$ interference users, see \cite{reddy2020structured} for more details. Let us call the first $m$ interference users as top interference users and last $n$ interference users as bottom interference users. Now, we calculate the number of colors in the closed anti-outneighborhood of Node (1,1) sequentially column by column. Due to symmetry, this number is same for all the users.  According to our placement policy, $\forall p\in[2:2m]$, colors of bottom interference nodes of a column $p$ matches with the colors of top interference nodes of its previous column. Therefore, they won't contribute to  closed anti-outneighborhood of Node (1,1). Therefore, the number of columns contributing to  closed anti-outneighborhood of Node (1,1) are $a_1+a_2+1$ in column 1 and $a_1+1$ if it is an odd column other than 1 and $a_2+1$ if it is an even column. Therefore, total number of colors in the closed anti-outneighborhood of Node (1,1) are $\lfloor \frac{K}{a_1+a_2+2} \rfloor (a_1+a_2+2)+a_2$. Therefore, $$\mathcal{X}_l\leq \lfloor \frac{K}{a_1+a_2+2} \rfloor (a_1+a_2+2)+a_2.$$
  
  Since, we are using only $K$ colors, $$\mathcal{X}_l\leq K.$$
  
  Therefore,
  $$\mathcal{X}_l\leq \min\{\lfloor \frac{K}{a_1+a_2+2} \rfloor (a_1+a_2+2)+a_2, K\}.$$ 
  
  Since, we divide each file into $\lfloor \frac{K}{a_1+a_2+2} \rfloor$ subfiles of equal size, 
  $$\overline{R}^*\leq \overline{R}_3=\frac{\min\{\lfloor \frac{K}{a_1+a_2+2} \rfloor (a_1+a_2+2)+a_2, K  \} }{\lfloor \frac{K}{a_1+a_2+2} \rfloor}.
  $$
 
\end{proof} 
  
  \section{Proof of the Section \ref{sec:results} theorems}\label{sec:proof}
  \begin{proof}
  We provide a proof sketch before giving the details. %According to our placement policy, we divide each file $\mathcal{F}_n$ $n\in [N]$ into $K$ disjoint subfiles $F_{n,m}$ $m\in[K]$  of equal size. Subfile $F_{n,m}$ is stored in caches $m, m+L, ..., m+(i-1)L$. Hence, Cache $k$ $k\in[K]$ stores the  subfiles $\mathcal{M}_k=\{\mathcal{F}_{n,<k-(r-1)L>_K}:n\in[N],r\in[i]\}$.
  We first store the files according to our placement policy.
  In the delivery phase, we form an instance of ICP with $K$ users, such that each user $j$ wants the unavailable $K-iL$ subfiles of File $\mathcal{F}_{d_j}$. The side-information of User $j$ is the subfiles needed for the remaining $K-1$ users and stored in user $j$'s available caches $[j:j+L-1]$. We show that the ICP formed is the union of several $\overline{(a_1,a_2)}_L-$ICPs and use the results in Section \ref{sec:ICP_preliminaries} to get an upper bound the optimal transmission rate of the ICPs. Finally, the optimal transmission rate of our MACC ICP is upper bounded by the sum of the upper bounds for the individual $\overline{(a_1,a_2)}_z-$ICPs. Different results in Section \ref{sec:ICP_preliminaries} gives different bounds. In particular, Theorem \ref{thm:newlocalchromatic} results give Theorem \ref{thm:BSR_UB_gen}, Lemma \ref{lem:BSR_UB} results give Theorem \ref{thm:our_linear_results} and Theorem \ref{thm:a1a2ICP_UB2} results give Theorem \ref{thm:our_nonlinear_results}. The details are as follows.
  
   %we form an instance of the ICP with the required subfiles and use the results mentioned in Section \ref{ICP_results} to derive an upper bound on the server transmission rate. In particular, we first split our ICP into many  ICP's of the form $\overline{(a_i,a_{i-1},...,a_1)}-$ICP. Then, we use Theorem \ref{thm:icpub} to get an upper bound on the data transmission rate of each $\overline{(a_i,a_{i-1},...,a_1)}-$ICPs. Finally, the data transmission rate of multi-access ICP is upper bounded by the sum of the upper bounds for the individual ICPs. The details are as follows.
   
   Recall that in the placement policy, we divide each file $\mathcal{F}_n$ $n\in [N]$ into $K$ disjoint subfiles $\mathcal{F}_{n,m}$ $m\in[K]$  of equal size. subfile $\mathcal{F}_{n,m}$ is stored in caches $m, m+L, ..., m+(i-1)L$. Hence, cache $k$ $k\in[K]$ stores the  subfiles $\mathcal{M}_k=\{\mathcal{F}_{n,<k-(r-1)L>_K}:n\in[N],r\in[i]\}$. 
   
   Some useful properties of the placement policy are
   \begin{itemize}
   	\item each cache stores $i$ subfiles of each file. Hence, the memory constraint $M=iN/K$ is satisfied. 
   	\item no two caches store the same set of subfiles
   	\item each subfile is available to exactly $iL$ consecutive users and no two subfiles are available to same set of caches
   	\item each user has access to $iL$ subfiles and no two users have same set of available subfiles
   \end{itemize} 
   
   Since, subfile $\mathcal{F}_{n,m}$ is stored in caches $m, m+L, ..., m+(i-1)L$, it is available to users with the indices in the set $[<m-L+1>_K:<m+(i-1)L>_K]$. Hence, $\mathcal{F}_{n,m}$ is represented as $\mathcal{F}_{n,[<m-L+1>_K:<m+(i-1)L>_K]}$, where the indices in the subscripted set indicates that the subfile available at corresponding users. For each subset of the set $[K]$ with $iL$ circular consecutive elements, there exists a subfile and vice versa. %Note that there are total $K$ subsets and $K$ subfiles.

   Let the request pattern be $\mathbf{d}=(d_1,d_2,...,d_K)\in[N]^K$, i.e., User $j$ requests File $\mathcal{F}_{d_j}$. In other words, User $j$ $j\in[K]$ requests $K$ subfiles of File $\mathcal{F}_{d_j}$. %According to our placement policy, we can represent every subfile of File $\mathcal{F}_n$ as $\mathcal{F}_{n,\mathcal{U}}$, where $\mathcal{U}$ is a subset of $[K]$ and it is a collection of $iL$ consecutive elements. For each $\mathcal{U}$, there exists a subfile. 
   Due to our placement policy, User $j$ already has access to $iL$ subfiles and it only needs remaining $K-iL$ subfiles of File $\mathcal{F}_{d_j}$.
   In particular, User $j$ already has access to $\mathcal{F}_{d_j,<j+l-1-(r-1)L>_K}$, $r\in[i]$,  $l\in[L]$ and needs the remaining $K-iL$ subfiles. The required subfiles at user $j$ are $\mathcal{F}_{d_j,[<j+a>_K:<j+a+iL-1>]}$, $\forall a\in [K-iL]$. By assuming worst case, each user requests a different file, a total of $K(K-iL)$ subfiles are needed across all the users.
   
   Now, we map our problem here to an instance of the ICP with $K$ users, such that each user $j$ wants the unavailable $K-iL$ subfiles of File $\mathcal{F}_{d_j}$. The side-information of User $j$ is the subfiles needed for the remaining $K-1$ users and stored in user $j$'s available caches $[j:j+L-1]$. To understand the structural properties of the ICP, we form a $K\times K-iL$ table such that
   \begin{enumerate}
   	\item each cell represents a virtual user,
   	\item $p^{\text{th}}$ row represents User $p$'s required subfiles,
   	\item Node ($p,q$) represents subfile  $\mathcal{F}_{d_p,[<p+q>_K:<p+q+iL-1>_K]}$.
   \end{enumerate}
 See Table \ref{Tab:gen_macc} for an illustration.

\begin{table}[h]
	%\begin{minipage}{.43\linewidth}
	\vspace{0.8cm}
	%\caption{}
	\centering
	%\resizebox{\linewidth}{!}{
		\begin{tabular}{| c | c | c | c | }
			\hline 
			$\mathcal{F}_{d_1,[2:iL+1]}$ &  $\mathcal{F}_{d_1,[3:iL+2]}$ & $\ldots$ &  $\mathcal{F}_{d_1,[K-iL+1:K]}$  \\
			\hline 
				$\mathcal{F}_{d_2,[3:iL+2]}$ &  $\mathcal{F}_{d_2,[4:iL+3]}$ & $\ldots$ &  $\mathcal{F}_{d_2,[K-iL+2:1]}$  \\
			\hline 
			\vdots & \vdots & \vdots & \vdots  \\
			\hline
				$\mathcal{F}_{d_K,[1:iL]}$ &  $\mathcal{F}_{d_K,[2:iL+1]}$ & $\ldots$ &  $\mathcal{F}_{d_K,[K-iL:K-1]}$  \\
			\hline 
		\end{tabular}
		\vspace{0.2cm}
		\caption{\sl The tabular representation of the  ICP formed by MACC problem. } \label{Tab:gen_macc}
	\end{table}	

Table \ref{Tab:gen_macc} represents a UICP with $K$ users each one requesting $K-iL$ subfiles. We convert the UICP into an SUICP as follows: in Table \ref{Tab:gen_macc}, each cell represents a subfile requested by the ICP user. Now, make each node as a virtual user in SUIP requesting corresponding subfile and the side-information is the same as the user represented by the row in UICP.

   Since all the nodes in a row are requested by the same user,    for all the nodes in a row the side-information nodes are same. In particular,  all the nodes which contain $l$ in the subscript are available at user $l$ and hence are side-information nodes to the nodes in Row $l$. Property 3 above implies that the relative positions of available subfiles are the same for all the nodes in a column and hence each column alone represents a symmetric ICP \cite{maleki2014index}. Note that (\emph{i}) the subscripted set of each subfile is a subset of $[K]$ with $iL$ consecutive elements and (\emph{ii}) in each column,  the subscripts are circularly shifting by one from Row $j$ to Row $j+1$. Thus for any column, each element $k \in [K]$ occurs in $iL$ consecutive entries. Hence, the side-information for any node in a column also follows the same pattern. %each element $l\in[K]$ occurs in $wL$ rows such that $w$ chunks each of size $L$  in each column and hence  the side-information within a column for any node occurs in $w$ chunks each of size $L$. 
   Therefore, each column represents a $(a_1,a_2)_{iL}-$ICP for some $a_1, a_2 \in \mathbb{N}$ and $a_1+a_2=K-iL-1$. %In particular, $b_{i+1}$ is the gap between a cell and its first side-information chunk, $b_{j}$ is the gap between the 
   In particular, column $j$ represents $(K-iL-j,j-1)_{iL}-$ICP. Therefore, according to the definition of $\overline{(a_1,a_2)}_z-$ICP, columns $j$ and $K-iL-j+1$ together represent $\overline{(K-iL-j,j-1)}_{iL}-$ICP $\forall j\leq \frac{K-iL}{2}$. Now, we use the results in Section \ref{sec:ICP_preliminaries} to upper bound the optimal transmission rate of each $\overline{(K-iL-j,j-1)}_{iL}-$ICP. The optimal transmission rate of the ICP in Table \ref{Tab:gen_macc} is upper bounded by the sum of the upper bounds for the individual $\overline{(K-iL-j,j-1)}_{iL}-$ICPs.
   
   Let $R_j^*$ be the optimal transmission rate of $\overline{(K-iL-j,j-1)}_{iL}-$ICP $\forall j\leq \frac{K-iL}{2}$, and $R_T^*$ be the optimal transmission rate of the ICP in Table \ref{Tab:gen_macc}. Then, $R_T^*$ is upper bounded as follows:
   \begin{itemize}
   	\item If $K-iL$ is even, then the ICP in Table \ref{Tab:gen_macc} can be represented as a union of exactly $\frac{K-iL}{2}$  $\overline{(a_1,a_2)}_z-$ICPs, given by $\overline{(K-iL-j,j-1)}_{iL}-$ICP $\forall j\leq \frac{K-iL}{2}$. Hence, 
   	\begin{equation}
   		R_T^*\leq \sum_{j=1}^{\frac{K-iL}{2}}R_j^*. \label{eq:even}
   	\end{equation}
   	\item If $K-iL$ is odd, then the ICP in Table \ref{Tab:gen_macc} can be represented as a union of  $\frac{K-iL-1}{2}$  $\overline{(a_1,a_2)}_z-$ICPs, given by $\overline{(K-iL-j,j-1)}_L-$ICP $\forall j\leq \frac{K-iL}{2}$ and one ${(a_1,a_2)}_z-$ICP given by ${(\frac{K-iL-1}{2},\frac{K-iL-1}{2})}_{iL}-$ICP. Hence, 
   	\begin{equation}
   		R_T^*\leq \sum_{j=1}^{\frac{K-iL-1}{2}}R_j^*+R_{\frac{K-iL-1}{2}}', \label{eq:odd}
   	\end{equation}
   	where $R_{\frac{K-iL-1}{2}}'$ is the optimal transmission rate of ${(\frac{K-iL-1}{2},\frac{K-iL-1}{2})}_{iL}-$ICP.
   \end{itemize}

Now, we prove Theorem \ref{thm:BSR_UB_gen}.

In a $\overline{(K-iL-j,j-1)}_{iL}-$ICP, $a_1=K-iL-j$ and $a_2=j-1$. Then, according to Theorem \ref{thm:newlocalchromatic}, if there exists an integer $X$ such that $X\geq a_1+a_2+2=K-iL+1$ and $X|K$, then $R_j^* \leq X$. Hence,
\begin{itemize}
	\item if $K-iL$ is even, then according to \eqref{eq:even}, we have 
	\begin{equation*}
		R_T^*\leq \sum_{j=1}^{\frac{K-iL}{2}}R_j^*=\frac{(K-iL)X}{2}. 
	\end{equation*}
\item if $K-iL$ is odd, then we can form an $\overline{(\frac{K-iL-1}{2},\frac{K-iL-1}{2})}_L-$ICP from ${(\frac{K-iL-1}{2},\frac{K-iL-1}{2})}_L-$ICP by splitting each file into 2 parts and show that ${(\frac{K-iL-1}{2},\frac{K-iL-1}{2})}_L-$ICP's optimal transmission rate is bounded by $X/2$, if there exists a $X$ such that $X\geq a_1+a_2+2=K-iL+1$ and $X|K$. Therefore, according to \eqref{eq:odd}, we have 
\begin{equation*}
R_T^*\leq \sum_{j=1}^{\frac{K-iL-1}{2}}R_j^*+R_{\frac{K-iL-1}{2}}=\frac{(K-iL-1)X}{2}+\frac{X}{2}. 
\end{equation*}
Note that, since we divide each subfile in ${(\frac{K-iL-1}{2},\frac{K-iL-1}{2})}_L-$ICP into 2 equal parts, the subpacketization for odd case is $K+1$, which is still linear.
\end{itemize}

Since, each subfile is of size $1/K$ units the optimal transmission rate of the MACC problem is upper bounded by $R_T^*/K$, this concludes the proof of Theorem \ref{thm:BSR_UB_gen}.

Now, we prove Theorem \ref{thm:our_linear_results}.

\begin{itemize}
	\item $(K-iL+1)|K$ case directly follows from Theorem \ref{thm:BSR_UB_gen}.
	\item $K-iL=1$ case forms an ICP such that each user want one subfile and has all the other users required subfiles. Hence, the side-information graph forms a clique and 1 message of size $1/K$ units is sufficient to solve the MACC problem.  
	\item  If $K-iL$ is even, then according to \eqref{eq:even}, we have
	\begin{equation*}
		R_T^*\leq \sum_{j=1}^{\frac{K-iL}{2}}R_j^*. 
	\end{equation*}
From Lemma \ref{lem:a1a2ICP_UB1}, we have $$R_j^*=\min\{a_1+2a_2+2,K\}=\min\{K-iL+j,K\}.$$ By substituting $R_j^*$ in  \eqref{eq:even}, with a little bit of algebraic manipulations, we get $ \frac{R_T^*}{K}	=R_{\text{e}}(M)$.
\item  If $K-iL$ is odd, then according to \eqref{eq:odd}, we have
\begin{equation*}
	R_T^*\leq \sum_{j=1}^{\frac{K-iL-1}{2}}R_j^*+R_{\frac{K-iL-1}{2}}'.
\end{equation*}
From Lemma \ref{lem:a1a2ICP_UB1}, we have $$R_j^*=\min\{a_1+2a_2+2,K\}=\min\{K-iL+j,K\},$$ and 
$$+R_{\frac{K-iL-1}{2}}'=K-iL.$$ By substituting $R_j^*$ and $+R_{\frac{K-iL-1}{2}}'$ in  \eqref{eq:odd}, with a little bit of algebraic manipulations, we get $ \frac{R_T^*}{K}	=R_{\text{o}}(M)$.
\end{itemize}
This completes the Theorem \ref{thm:our_linear_results} proof.

Similarly, if we use Theorem \ref{thm:newlocalchromatic} results to bound the transmission rate ($R_T^*$) of MACC ICP in Table \ref{Tab:gen_macc}, we get Theorem \ref{thm:our_nonlinear_results} upper bound. Since, we divide each file into $a_1+a_2+2$ subfiles to get \ref{thm:newlocalchromatic}, the subpacketization is $\widetilde{F}_5(M).$ 

\end{proof}

\section{Example}

The following example illustrates our policy.

\begin{example}\label{ex:detailed_example}
	Consider a MACC problem with $N=K=8$, $L=2$ and $M=i=3$.\\
	\textit{Placement policy:}
	\begin{itemize}
		\item Divide each file $\mathcal{F}_n$ $\forall n\in[8]$  into $K=8$ subfiles $\mathcal{F}_{n,1},\mathcal{F}_{n,2},\mathcal{F}_{n,3},\mathcal{F}_{n,4},\mathcal{F}_{n,5},\mathcal{F}_{n,6}$, $\mathcal{F}_{n,7}$ and $\mathcal{F}_{n,8}$.
		\item Cache $k$ stores $i=3$ subfiles of each file which are given by $\mathcal{F}_{n,k}, \mathcal{F}_{n,<k-2>_8}$ and $\mathcal{F}_{n,<k-4>_8},$ $\forall n\in[8]$.
	\end{itemize}
	Note that according to our placement policy, $\mathcal{F}_{n,m}$ $\forall n,m\in[8]$ is stored in caches $m$, $<m+2>_8$ and $<m+4>_8$ and is  available to users in the set $[<m-1>_8:<m+4>_8]$, because users $<m-1>_8,m$ are connected to cache $m$, users $<m+1>_8,$, $<m+2>_8$ are connected to cache $<m+2>_8$ and users $<m+2>_8,$, $<m+4>_8$ are connected to cache $<m+3>_8$. Hence, $\mathcal{F}_{n,m}$ can also be represented as $\mathcal{F}_{n,[<m-1>_8:<m+4>_8]}$, where the subscripted set $[<m-1>_8:<m+4>_8]$ indicates that subfile $\mathcal{F}_{n,[<m-1>_8:<m+4>_8]}$ is available at users in the set $[<m-1>_8:<m+4>_8]$. Note that the set $[<m-1>_8:<m+4>_8]$ is a collection of $iL=6$ circular consecutive{\footnote{In our paper, circular consecutive means consecutive with cyclic wrap-around.}} elements\\
	
	\textit{Delivery policy:}
	\begin{itemize}
		\item Let the request pattern be $(d_1,d_2,d_3,d_4,d_5,d_6, d_7, d_8)$. 
		\item Each user $j$ wants 8 subfiles of File $\mathcal{F}_{d_j}$ and it already has access to $iL=6$ subfiles. Hence, User $j$ needs 2 subfiles of  File $\mathcal{F}_{d_j}$ given by $\mathcal{F}_{d_j,\{<j+1>_8, <j+2>_8, <j+3>_8, <j+4>_8,<j+5>_8,<j+6>_8\}}$,  and $\mathcal{F}_{d_j,\{<j+2>_8, <j+3>_8, <j+4>_8,<j+5>_8,<j+6>_8, <j+7>_8\}}$, i.e.,
		\iffalse	\begin{itemize}
			\item User 1 needs $\mathcal{F}_{d_1,5}$, $\mathcal{F}_{d_1,6}$, $\mathcal{F}_{d_1,7}$ and $\mathcal{F}_{d_1,8}$,
			\item User 2 needs $\mathcal{F}_{d_2,6}$, $\mathcal{F}_{d_2,7}$, $\mathcal{F}_{d_2,8}$ and $\mathcal{F}_{d_2,1}$,
			\item User 3 needs $\mathcal{F}_{d_3,7}$, $\mathcal{F}_{d_3,8}$, $\mathcal{F}_{d_3,1}$ and $\mathcal{F}_{d_3,2}$,
			\item User 4 needs $\mathcal{F}_{d_4,8}$, $\mathcal{F}_{d_4,1}$, $\mathcal{F}_{d_4,2}$ and $\mathcal{F}_{d_4,3}$,
			\item User 5 needs $\mathcal{F}_{d_5,1}$, $\mathcal{F}_{d_5,2}$, $\mathcal{F}_{d_5,3}$ and $\mathcal{F}_{d_5,4}$,
			\item User 6 needs $\mathcal{F}_{d_6,2}$, $\mathcal{F}_{d_6,3}$, $\mathcal{F}_{d_6,4}$ and $\mathcal{F}_{d_6,5}$,
			\item User 7 needs $\mathcal{F}_{d_7,3}$, $\mathcal{F}_{d_7,4}$, $\mathcal{F}_{d_7,5}$ and $\mathcal{F}_{d_7,6}$, and
			\item User 8 needs $\mathcal{F}_{d_8,4}$, $\mathcal{F}_{d_8,5}$, $\mathcal{F}_{d_8,6}$ and $\mathcal{F}_{d_8,7}$.
		\end{itemize}
		In other words, \fi
		\begin{itemize}
			\item User 1 needs $\mathcal{F}_{d_1,\{2,3,4,5,6,7\}}$, and $\mathcal{F}_{d_1,\{3,4,5,6,7,8\}}$,
			\item User 2 needs $\mathcal{F}_{d_2,\{3,4,5,6,7,8\}}$,  and $\mathcal{F}_{d_2,\{4,5,6,7,8,1\}}$,
			\item User 3 needs $\mathcal{F}_{d_3,\{4,5,6,7,8,1\}}$,  and $\mathcal{F}_{d_3,\{5,6,7,8,1,2\}}$,
			\item User 4 needs $\mathcal{F}_{d_4,\{5,6,7,8,1,2\}}$,  and $\mathcal{F}_{d_4,\{6,7,8,1,2,3\}}$,
			\item User 5 needs $\mathcal{F}_{d_5,\{6,7,8,1,2,3\}}$,  and $\mathcal{F}_{d_5,\{7,8,1,2,3,4\}}$,
			\item User 6 needs $\mathcal{F}_{d_6,\{7,8,1,2,3,4\}}$,  and $\mathcal{F}_{d_6,\{8,1,2,3,4,5\}}$,
			\item User 7 needs $\mathcal{F}_{d_7,\{8,1,2,3,4,5\}}$,  and $\mathcal{F}_{d_7,\{1,2,3,4,5,6\}}$, and
			\item User 8 needs $\mathcal{F}_{d_8,\{1,2,3,4,5,6\}}$,  and $\mathcal{F}_{d_8,\{2,3,4,5,6,7\}}$.
		\end{itemize}
		\item Now we map the coded caching problem to an ICP and represent the ICP in a tabular form as follows:
		
			\begin{table}[h]
			%	\vspace{-0.25in}
			%\begin{minipage}{.6\textwidth}
			%\caption{}
			\centering
			%\resizebox{\linewidth}{!}{
				\begin{tabular}{| c | c |  }
					\hline 
					\color{red}	$\mathcal{F}_{d_1,\{2,3,4,5,6,7\}}$ & \color{red} $\mathcal{F}_{d_1,\{3,4,5,6,7,8\}}$  \\
					\hline 
					$\mathcal{F}_{d_2,\{3,4,5,6,7,8\}}$ & \color{blue} $\mathcal{F}_{d_2,\{4,5,6,7,8,1\}}$  \\
					\hline
					\color{blue}	$\mathcal{F}_{d_3,\{4,5,6,7,8,1\}}$ & \color{blue} $\mathcal{F}_{d_3,\{5,6,7,8,1,2\}}$  \\
					\hline 
					\color{blue}	$\mathcal{F}_{d_4,\{5,6,7,8,1,2\}}$ & \color{blue} $\mathcal{F}_{d_4,\{6,7,8,1,2,3\}}$  \\
					\hline
					\color{blue} $\mathcal{F}_{d_5,\{6,7,8,1,2,3\}}$ & \color{blue} $\mathcal{F}_{d_5,\{7,8,1,2,3,4\}}$\\
					\hline
					\color{blue} 	$\mathcal{F}_{d_6,\{7,8,1,2,3,4\}}$ & \color{blue} $\mathcal{F}_{d_6,\{8,1,2,3,4,5\}}$ \\
					\hline
					\color{blue}	$\mathcal{F}_{d_7,\{8,1,2,3,4,5\}}$ & \color{blue} $\mathcal{F}_{d_7,\{1,2,3,4,5,6\}}$\\
					\hline
					\color{blue}	$\mathcal{F}_{d_8,\{1,2,3,4,5,6\}}$ & $\mathcal{F}_{d_8,\{2,3,4,5,6,7\}}$  \\
					\hline 
					\hline 
					$(1,0)_6-$ICP & $(0,1)_6-$ICP  \\
					\hline 
				\end{tabular}
				\vspace{0.2cm}
				\caption{ICP for $(N=8,K=8)$-CCDN. Row $p$ contains the subfiles needed for User $p$, $\forall p\in[8]$. In the last row, we mentioned the ICP type formed by the nodes of the corresponding column. We highlight User 1's needed sub-files with red color and their side-information sub-files with blue color. Note that all the blue colored sub-files contain 1 in their subscripted set. Hence, the blue colored nodes are side-information nodes for Row 1 nodes.}  \label{Tab:881}
				%	\end{minipage}%
			\hspace{0.01\linewidth}
			
		\end{table}
		
		Total 16 subfiles are involved in the server transmission. We can map the problem here to an instance of the index coding problem, with $K=16$ virtual users, each one requesting a distinct subfile. The side information at the  virtual user representing (and requesting) some Subfile $X$ are the subfiles available to the  user which is requesting Subfile $X$. For example, the side information of the  virtual user representing Subfile $\mathcal{F}_{d_1,\{2,3,4,5,6,7\}}$ are the subfiles available to User 1, i.e., $\mathcal{F}_{d_2,\{4,5,6,7,8,1\}}$, $\mathcal{F}_{d_3,\{5,6,7,8,1,2\}}$, $\mathcal{F}_{d_3,\{6,7,8,1,2,3\}}$, ... . 
		We can solve this index coding problem to get the achievable transmission rate for our proposed scheme. Note that the number of messages transmitted by the central server in the CCDN is equal to the number of messages transmitted in the ICP, and the size of each message is equal to the size of a  subfile.

		To understand the structural properties of the above ICP, we form a $8\times2$ table (see Table \ref{Tab:881}), such that the $p^{\text{th}}$ row and $q^{\text{th}}$ column contains sub-file ${F}_{d_p,[<p+q>_K:<p+q+5>_K]}$, i.e., the subfile requested by User $p$ and available at users $<p+q>,$ $<p+q+1>$, $<p+q+2>$, $<p+q+4>_K$, $<p+q+5>_K$ and $<p+q+6>$.  We refer to this entry as the Node $(p,q)$  where, $p\in[8],q\in[2]$. Note that the entries in Row $p$ are the subfiles needed for User $p$.
		%We will use the notation $(p,\bar{q})$ to represent all the other nodes in Row $p$ excluding Node $(p,{q})$. For our example $(p,\bar{1})$ represents Node $(p,2)$ and $(p,\bar{2})$ represents Node $(p,1)$. 

		%\end{itemize}
		
		Hence, we map the coded caching problem to an ICP and the ICP is represented in a tabular form in Table \ref{Tab:881}. The ICP formed is symmetrical ICP, which is defined as if Node ($p_1,q_1$) is a side-information (or interference) node of Node ($p_2,q_2$), then Node ($<p_1+1>_K,q_1$) is a side-information (or interference) node of Node ($<p_2+1>_K,q_2$) and vice versa. To illustrate that the ICP formed is symmetric, we represent the ICP corresponding  to  $(N=8,K=8)-$CCDN in Tables \ref{Tab:881} and \ref{Tab:882} such that  Table \ref{Tab:881} highlights User 1's requested files and its side-information files and Table \ref{Tab:882} highlights User 2's requested files and its side-information files. Observe that in the tables' Column 1, User 1 needs Sub-file $\mathcal{F}_{d_1,\{2,3,4,5,6,7\}}$ and followed by it there are 1 interference node and after that 6 side-information nodes of User 1. This is highlighted in Table \ref{Tab:881}.  Similarly,  User 2 needs Sub-file $\mathcal{F}_{d_2,\{3,4,5,6,6,7\}}$ and followed by it there are 1 interference node and after that 6 side-information nodes of User 2. This is highlighted in Table \ref{Tab:882}. Due to symmetry, in Column 1 of the tables, User $i$ requests $\mathcal{F}_{d_i,[<i+1>_8:<i+6>_8]}$ and followed by it there are 1 interference nodes and after that 6 side-information nodes of User $i$. Hence, Column 1 represents $(a_1=1,a_2=0)_6-$ICP. Similarly, we can also check that in Column 2 of the tables, User $i$ requests $\mathcal{F}_{d_i,[<i+2>_8:<i+7>_8]}$ and followed by it there are 6 side-information nodes, and after that 1 interference node of User $i$. Hence, Column 2 represents $(a_1=0,a_2=1)_6-$ICP. Union of these 2 ICPs is  $\overline{(a_1=1,a_2=0)}_6-$ICP. Hence, Tables \ref{Tab:881} and \ref{Tab:882} represent $\overline{(a_1=1,a_2=0)}_6-$ICP.

		\begin{table}[h]
			%	\vspace{-0.25in}
			%\begin{minipage}{.6\textwidth}
			%\caption{}
			\centering
			%\resizebox{\linewidth}{!}{
				\begin{tabular}{| c | c |  }
					\hline 
					\color{blue}	$\mathcal{F}_{d_1,\{2,3,4,5,6,7\}}$ &  $\mathcal{F}_{d_1,\{3,4,5,6,7,8\}}$  \\
					\hline 
					\color{red}	$\mathcal{F}_{d_2,\{3,4,5,6,7,8\}}$ & \color{red} $\mathcal{F}_{d_2,\{4,5,6,7,8,1\}}$  \\
					\hline
					$\mathcal{F}_{d_3,\{4,5,6,7,8,1\}}$ & \color{blue} $\mathcal{F}_{d_3,\{5,6,7,8,1,2\}}$  \\
					\hline 
					\color{blue}	$\mathcal{F}_{d_4,\{5,6,7,8,1,2\}}$ & \color{blue} $\mathcal{F}_{d_4,\{6,7,8,1,2,3\}}$  \\
					\hline
					\color{blue} $\mathcal{F}_{d_5,\{6,7,8,1,2,3\}}$ & \color{blue} $\mathcal{F}_{d_5,\{7,8,1,2,3,4\}}$\\
					\hline
					\color{blue} 	$\mathcal{F}_{d_6,\{7,8,1,2,3,4\}}$ & \color{blue} $\mathcal{F}_{d_6,\{8,1,2,3,4,5\}}$ \\
					\hline
					\color{blue}	$\mathcal{F}_{d_7,\{8,1,2,3,4,5\}}$ & \color{blue} $\mathcal{F}_{d_7,\{1,2,3,4,5,6\}}$\\
					\hline
					\color{blue}	$\mathcal{F}_{d_8,\{1,2,3,4,5,6\}}$ & \color{blue} $\mathcal{F}_{d_8,\{2,3,4,5,6,7\}}$  \\
					\hline 
					\hline 
					$(1,0)_6-$ICP & $(0,1)_6-$ICP  \\
					\hline 
				\end{tabular}
				\vspace{0.2cm}
				\caption{ICP for $(N=8,K=8)$-CCDN.  We highlight User 2's needed sub-files with red color and their side-information sub-files with blue color. Note that all the blue colored sub-files contain 2 in their subscripted set. Hence, the blue colored nodes are side-information nodes for Row 2 nodes.}  \label{Tab:882}
				%	\end{minipage}%
			\hspace{0.01\linewidth}
			
		\end{table}
		
		From, Lemma \ref{lem:a1a2ICP_UB1}, $\bar{R}_1=3$ and 3 messages are enough to solve the given MACC problem. Since, each subfile is of size $1/K$ units, our transmission rate is $3/8$ units. Hence, our transmission rate is $3/8$ units and subpacketization $F=K=8$, whereas from Lemma \ref{lem:chinese_ub}, transmission rate  $R_2(M)=0.5 $ units and subpacketization $F=4K=32$.
		
	\end{itemize}		
\end{example}